\documentstyle[multicol,aps,prl,epsf,array]{revtex}
\begin{document}
\title{Subsystem purity as an enforcer of entanglement}
\author{S. Bose,$^{\dagger}$ I. Fuentes-Guridi,$^{*}$ P.L.Knight,$^{*}$ and V. Vedral$^{*}$}
\address{$^{*}$Optics Section, The Blackett Laboratory,
Imperial College, London SW7 2BW, United Kingdom \\
$^{\dagger}$Centre for Quantum Computation, Clarendon Laboratory,
    University of Oxford,
    Parks Road,
    Oxford OX1 3PU, England}

\maketitle
\begin{abstract}
We show that entanglement can always arise in the interaction of
an arbitrarily large system in any mixed state with a single qubit
in a pure state. This small initial purity is enough to enforce
entanglement even when the total entropy is close to maximum. We
demonstrate this feature using the Jaynes-Cummings interaction of
a two level atom in a pure state with a field in a thermal state
at an arbitrarily high temperature. We find the time and
temperature variation of a lower bound on the amount of
entanglement produced and study the classical correlations
quantified by the mutual information.
\end{abstract}


\begin{multicols}{2}

 Entanglement is a key resource in quantum information
processing \cite{revs1,revs2}. In practice, due to decoherence, it
is very difficult to generate and maintain entanglement in the
form of a pure state. With this motivation, there has been a lot
of recent interest in understanding and quantifying entanglement
of mixed states \cite{vedral97,bennett96,wootters98}. As
entanglement is an important resource, we need to investigate
whether and how it can be generated in severe conditions of
mixedness \cite{vedral97}. Interestingly, if we had a qubit in a
completely pure state $|0\rangle_1$ and another qubit in a
completely mixed state $(1/2)(|0\rangle\langle 0|+|1\rangle\langle
1|)_2$ , we could let them interact through
\begin{eqnarray}
|0\rangle_1|0\rangle_2 &\rightarrow& |0\rangle_1|0\rangle_2 \nonumber \\
|0\rangle_1|1\rangle_2 &\rightarrow&  |\psi^{+}\rangle_{12},
\end{eqnarray}
where
$|\psi^{+}\rangle_{12}=\frac{1}{2}(|01\rangle_{12}+|10\rangle_{12})$,
to generate an entangled state. Surprisingly, this also holds for
a pure qubit interacting with a fully mixed state of an arbitrary
dimensional (even macroscopic) system. In this letter, we
demonstrate this in the context of a pure state atom interacting
with mixed state quantum field (an infinite dimensional system)
through the Jaynes-Cummings model.

         The Jaynes Cummings model (JCM)
\cite{original,knight} is the simplest model that describes the
interaction between light and matter. It consists of a two level
atom interacting with a near-resonant quantized mode of the field.
The atom and field become dynamically entangled by their
interaction. This model provides direct evidence for the quantum
mechanical nature of the electromagnetic field by predicting
collapses and revivals of Rabi oscillations of the atom
\cite{collapse}. These have been tested experimentally
\cite{morejcm,haroche}. The JCM has analytical solutions for
arbitrary coupling constants and can easily be extended to include
a variety of initial conditions \cite{haroche2}, dissipation and
damping \cite{barnett}, multiple atoms \cite{travis}, multilevel
atoms \cite{abdel}, multimode fields \cite{parker} and more
elaborate interactions \cite{knight2}. For certain pure initial
states of the atom and the cavity mode, the JCM produces
entanglement which oscillates with time \cite{Phoenix}. In this
paper we show that even if the cavity field is initially in a {\em
thermal state}, entanglement {\em still} arises in course of the
JCM interaction. The result holds {\em irrespective of the
temperature of the field}. This goes counter to the folklore that
entanglement, being a very quantum attribute, should automatically
disappear at high temperatures. We give a lower bound on the
entanglement produced as a function of the temperature of the
cavity field and time. We also compare the entanglement, which
quantifies the quantum part of the correlations, to the total
correlations as quantified by the mutual information.

Entanglement is well understood for pure states of bipartite
systems such as the JCM. The general state of a system involving two
subsystems (say atom and field) can be written as a superposition
of the products of individual states. If the state is pure it can
always be written in the Schmidt form \cite{ekert}
$$
\vert\Psi\rangle=\sum_{n}g_{n}\vert u_{n}^{\prime}\rangle_a\vert
v_{n}^{\prime}\rangle_f
$$
where $\vert u_{n}^{\prime}\rangle_a$ and  $\vert
u_{n}^{\prime}\rangle_f$ are orthonormal bases for the atomic and
field subsystems respectively. The correlations of the two systems
are then fully displayed. In this case (pure bipartite states),
entanglement is quantified by the entropy of the reduced density
matrix of either of the subsystems defined as
$S(\rho)=-Tr(\rho_{a}\log(\rho_{a}))$. This quantification of
entanglement has been used in all earlier studies of entanglement
in JCM with pure initial states \cite{Phoenix}. For the JCM with
both pure atomic and cavity field states, the cavity field can be
considered as an effective two level system \cite{Phoenix}.

 Here we are interested in studying entanglement for an initial
mixed state of the cavity field, as mixed states are the true
representation of the state of the field at a finite temperature.
Entanglement for mixed states is difficult to define. This is
because we cannot easily define an analogue of the Schmidt
decomposition for a general mixed state of a composite system.
Such a mixed state can be expanded in terms of pure states in
infinitely many different ways and it is not clear which, if any
decomposition should be favored. Mixing two entangled pure states
could result in a mixed state with entanglement much less than the
average entanglement of the states mixed. Mixed state entanglement
is thus a very different entity to either correlations or pure
state entanglement. At least three different measures have been
used to quantify entanglement for a mixed state. One of these
measures, the relative entropy of entanglement \cite{vedral97}, is
defined for mixed state $\rho_{af}$ of a composite system (such as
the atom-field system in the JCM) as
$$
E_{re}(\rho_{af})=\min_{\sigma\in\em{D}}(Tr(\rho_{af}(\log\rho_{af}-\log\sigma_{af}))),
$$
where $\em{D}$ is the set of disentangled (separable) states of
the system. A disentangled state can be written in the form
$\sum_i p_i \rho_a^i\otimes\rho_f^i$. This measure tells us how
difficult it is to distinguish the given entangled state from its
closest approximation to the set of separable states. The other
measures of entanglement are associated with formation and
distillation of entangled states. Consider the number, n, of
copies of a non maximally entangled state $\rho_{af}$ that can be
created by using only {\em correlated local actions} (i.e. through
correlated actions on the field state and atomic state alone) on a
number, m, of maximally entangled states. Entanglement of
formation is the asymptotic conversion ratio, $m/n$ in the limit
of infinitely many copies \cite{wootters},
$$
E_{f}(\rho_{af})=\min\sum_{i}p_{i}S(\rho_{a}^{i})
$$
where the minimum is taken over all the possible realizations of
the state $\rho_{af}=\sum_{i}p_{i}\vert\Psi_{af}^{i}\rangle\langle
\Psi_{af}^{i}\vert$. A measure stemming from the opposite process
(distillation) is the entanglement of distillation. It is the
asymptotic rate $m/n$ of converting $m$ copies of a non-maximally
entangled state $\rho_{af}$ into $n$ copies of a maximally
entangled state by means of correlated local actions. The
entanglement of distillation is in general smaller than formation.
All the different measures of entanglement are related to each
other through the amount of available classical information about
the decomposition of the state \cite{leah}.

  It is not easy to
compute the value of entanglement from the measures. The
entanglement of formation is the only measure for which an
analytical method exists for calculating the entanglement, but
this is specific to the case a state of $2\times2$ systems.
However, in our case there is a two level atom interacting with a
cavity field, which is an infinite dimensional system.  For such
states, we can still give a lower bound on the entanglement from
the known result in the case of $2\times2$ systems. We first
project the entire atom-field state onto a subspace equivalent to
a $2\times2$ system. We can then compute the entanglement of
formation for each of the outcomes. This particular projection
onto a $2\times2$ system, as we will show, can be done by local
actions alone. Being local, such an action cannot increase the
entanglement on average \cite{vedral97}.  If we compute the
average of the entanglement over all possible outcomes, the result
will thus be a lower bound on the entanglement in the initial
$2\times \infty$ state of the atom and the cavity field. We will
also look at the total correlations between the atom and the mixed
field as quantified by the mutual information \cite{ingarden},
$$
I=S(\rho_{a})+S(\rho_{f})-S(\rho_{af})
$$
where $\rho_{a}$ and $\rho_{f}$ are the reduced density matrices
of the atom and field. This measure quantifies how much the
correlated systems know about the state of each other. As the
mutual information indicates the total correlations, it should be
larger than the lower bound on entanglement we compute. We will
compare the two quantities (i.e. the lower bound on the
entanglement and the mutual information) to understand how much
the purely quantum correlations contribute to the total
correlations.

We consider the field in our JCM example initially to be in a
thermal state at some temperature $T$ with the probability
distribution $P_{n}$ for number states $|n\rangle$ being given by
\begin{equation} \label{eq:initial}
P_{n}=\frac{1}{1+\langle n\rangle}\left(\frac{\langle n\rangle}{1+\langle n\rangle}\right)^{n}
\end{equation}
where $\langle n \rangle={\{e^{\beta\hbar\omega}-1\}}^{-1}$ is
the mean photon number, $\beta=1/k_{B}T$ and $k_{B}$ is
Boltzmann's constant, $T$ is the temperature, $\omega$ the
frequency of the optical mode and $\hbar$ the Planck's constant.
The two-level atom is initially taken to be in the excited state
$|e\rangle_a$ (the ground state being $|g\rangle_a$). The JCM
interaction between the atom and the field is given by
\begin{equation}
H_{\mbox{JCM}}=g(\vert e\rangle\langle g\vert_a a_f+ a^{\dagger}_f
\vert g\rangle \langle e\vert_a),
\end{equation}
where $a_f$ and $a^{\dagger}_f$ are the annihilation and creation
operators of the field mode respectively.  The joint density
matrix for the atom-field system evolves with time as

\begin{equation}
\rho_{af}=\sum_{n=0}^{\infty}P_{n}\rho_{n},
\label{net}
\end{equation}
where
\begin{eqnarray}
\rho_{n}&=&\cos^{2}{\left(\frac{\Omega_{n}t}{2}\right)}\vert e,n\rangle\langle e,n\vert_{af}\nonumber\\
&-&i\cos{\left(\frac{\Omega_{n}t}{2}\right)}\sin{\left(\frac{\Omega_{n}t}{2}\right)}\vert e,n\rangle\langle g,n+1\vert_{af} \nonumber\\
&+&i\cos{\left(\frac{\Omega_{n}t}{2}\right)}\sin{\left(\frac{\Omega_{n}t}{2}
\right)}\vert g,n+1\rangle\langle e,n\vert_{af} \nonumber\\
&+&\sin^{2}{\left(\frac{\Omega_{n}t}{2}\right)}\vert
g,n+1\rangle\langle g,n+1\vert_{af}
\end{eqnarray}
where $\Omega_{n}=2g\sqrt{n+1}$ is the Rabi frequency.

\begin{figure}
\leavevmode \epsfxsize=8cm \epsfbox{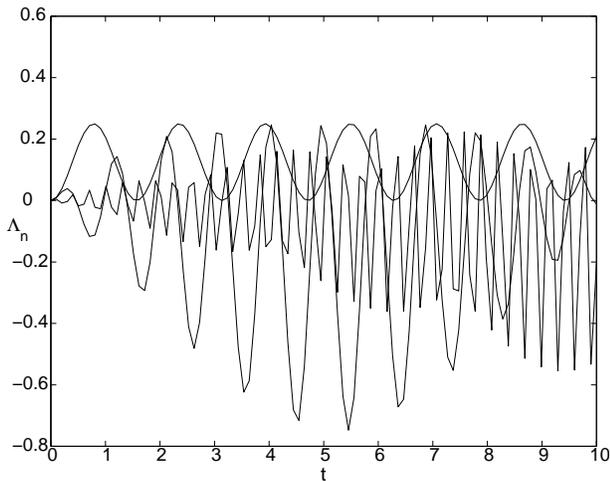} \caption{Plots
of the time variation of the inseparability expression $\Lambda$
for three values of $n$. The fastest oscillating curve is for
$n=100$, the next fastest for $n=10$ and the slowest oscillating
curve is for $n=0$.} \label{ineq}
\end{figure}

 We cannot exactly compute the total atom-field entanglement in the mixed state
 given by the above Eq.(\ref{net}). However, we can obtain an estimate a lower bound on the entanglement using the
 projection to a $2\times 2$ subspace discussed earlier.
To see this, consider the field state being projected into the
subspace spanned by $|n\rangle_f$ and $|n+1\rangle_f$. As this is
a local action on the field, it cannot increase the entanglement.
The resulting state is
\begin{displaymath}
\rho_{af}^{n}= \left(\begin{array}{cccc}
P_{n-1}S_{n-1}^{2} & 0 & 0 & 0 \\
0 & P_{n}S_{n}^{2} & P_{n}iC_{n}S_{n} & 0 \\
0 & P_{n}iC_{n}S_{n} & P_{n}C_{n}^{2} & 0 \\
0 & 0 & 0 & P_{n+1}C_{n+1}^{2}\end{array}\right)
\end{displaymath}
where
\begin{equation}
C_{n}=\cos{\left(\frac{\Omega_{n}t}{2}\right)},
~S_{n}=\sin{\left(\frac{\Omega_{n}t}{2}\right)}.
\end{equation}
Before proceeding to the evaluation of entanglement based on the
above formula, we would check the separability of the above state.
To prove the inseparability of this matrix we compute the
eigenvalues of the partially transposed matrix. If one of the
eigenvalues is negative then $\rho_{af}^{n}$ is inseparable
\cite{horo}. The existence of the negative eigenvalue reduces to
the condition
\begin{eqnarray}
\label{expr1}
(P_{n}C_{n}S_{n})^{2}>P_{n-1}P_{n+1}(C_{n+1}S_{n-1})^{2}.
\end{eqnarray}
Substituting $P_n, P_{n+1}$ and $P_{n-1}$ from
Eq.(\ref{eq:initial}) in the above expression we obtain the
condition
\begin{eqnarray}
\label{expr2}
 \Lambda_n=(C_{n}S_{n})^{2}-(C_{n+1}S_{n-1})^{2}>0,
\end{eqnarray}
which is {\em independent} of $\langle n\rangle$ (i.e. of the
temperature). We plot the expression $\Lambda_n$ with time $t$ in
Fig.\ref{ineq} for three values of $n$.

From Fig.\ref{ineq}, we see that the expression $\Lambda_n$ is
positive for some intervals of time for each $n$ implying
$\rho_{af}^{n}$ is entangled during those intervals of time. The
entire atom-cavity state $\rho_{af}$ is thus also entangled in
those intervals of time (otherwise, no local projection would have
given an entangled outcome). Entanglement thus arises due to JCM
interaction of an atom and a field in a thermal state {\em
irrespective} of the temperature of the field. Note that this is a
rigorous result because even if $\Lambda$ is negative for just one
specific value of $(n,t)$, then a local projection at that $t$
will result in the entangled outcome $\rho_{af}^{n}$ with a finite
probability. This would mean that prior to the local measurement,
entanglement was present in the atom-field state at that time.

  Based on the plots in Fig.\ref{ineq}, we heuristically justify
the following conjecture: {\em Entanglement is present at all
times except at $t=0$}. The conjecture relies on the observed
behaviour that for higher values of $n$, oscillations of
$\Lambda_n$ (separability) with $t$ are faster. Consider an
arbitrarily small interval of time $\delta t < \epsilon$. By
going to sufficiently high $n$, one can always find that
$\Lambda_n$ has a period smaller than $\delta t$ and thereby
$\rho_{af}^{n}$ is entangled for an interval of time within
$\delta t$. We can choose the time interval $\delta t$ smaller and
smaller, so that eventually there is entanglement at all instants
of time.

We now briefly comment on the alternative scenario where the atom
is in the initial thermal state
\begin{equation}
\lambda|e\rangle\langle e|+(1-\lambda)|g\rangle\langle g|,
\end{equation}
where $\lambda/(1-\lambda)=\exp(-\Delta E/kT)$ ($\Delta E$ being
the energy difference between $|e\rangle$ and $|g\rangle$ and $T$
the temperature). The field is assumed to start in a pure Fock
state $|n\rangle$. As the JCM is completely {\em symmetric}
between the atomic and field operators, we can exchange the field
and the atom states and have the same entanglement. This new
scenario is equivalent to just two levels of the field involved in
interaction with a pure atom. As shown earlier this also leads to
entanglement at all temperatures of the atom.

The case when both the atom and the field are in thermal states is
much more complex. In this case we expect that there is a cut-off
temperature above which there is no entanglement. At infinite
temperature, of course, all entanglement disappears as the total
state is just proportional to the identity. However, our
projection method can only lead to lower bounds as the failure of
our method to produce entanglement by local projections does not
imply that entanglement does not exist in the original mixed
state. So, until there an operational necessary and sufficient
condition for separability of $2\times N$ density matrices, we
cannot fully address this case.
\begin{figure}
\leavevmode \epsfxsize=8cm \epsfbox{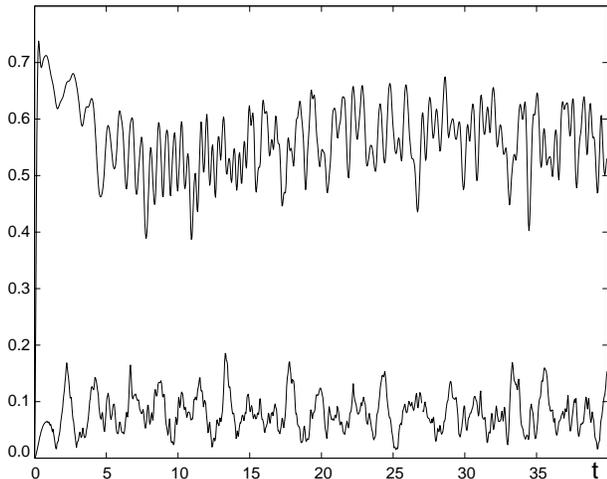} \caption{Plots of
the time variation of the mutual information and the entanglement
in a thermal JCM model with $\langle n \rangle=10$. The upper
curve represents mutual information and the lower curve represents
entanglement.} \label{figure1}
\end{figure}
As mentioned earlier, if we evaluate the entanglement of
$\rho_{af}^n$ and take an average over all possible values of $n$
(weighted by the probability of obtaining $\rho_{af}^n$), we
should get a lower bound on the entanglement of the state
$\rho_{af}$. This holds because $\rho_{af}^n$ is the result of a
local projection on $\rho_{af}$. We have computed this lower bound
on entanglement and the mutual information  for initial thermal
states with various mean photon numbers (e.g. a plot for the mean
photon number $\langle n\rangle=10$ is shown in Fig. 2). We find,
as we proved earlier, that JCM generates entanglement at {\em any
temperature} of the cavity field, no matter how high. Moreover,
though it starts from zero, it never completely vanishes. This
supports our earlier conjecture that entanglement is present at
all times except at $t=0$. Mutual information, on the other hand,
quantifies total correlations and is therefore greater or equal to
entanglement as seen in Fig. 2. Classical correlations are, of
course, always seen to be present and this means that the total
state is never of the product form at $t>0$ due to the JCM
interaction. The mutual information is a good measure of classical
correlations when entanglement is absent, since then total
correlations are equal to classical correlations.

We have shown that entanglement can be generated from a very small
amount of purity existing in an overall very mixed state. In
principle, a $2\times N$ system can be entangled even if the total
entropy is as high as $\log N$ (the maximum being $1+\log N$).
This is an important result among attempts to relate mixedness and
entanglement \cite{mix}. This means in principle, one could
entangle a microscopic system in a pure state with a macroscopic
system in a thermal state, as suggested by the results in
\cite{Jacobs}. Our results should have implications for quantum
computation with mixed states \cite{Luke}. This kind of
entanglement is probably involved in the functioning of Shor's
algorithm with only one pure qubit \cite{Mosca}. Further studies
could involve other natural interactions. For example, one could
study how the creation of entanglement is affected by the off
resonant interaction of the atom and the field. It is also an
interesting and open question how efficiently one can entangle two
partially mixed systems in general.

This research has been partly supported by the European Union,
The UK Engineering and Physical Sciences Research Council and
Hewlett-Packard. I. F.-G. would like to thank Consejo Nacional de
Ciencia y Tecnologia (Mexico) Grant no. 115569/135963 for
financial support.


\end{multicols}{2}

\end{document}